\RequirePackage{fix-cm}
\documentclass[twocolumn,epjc3]{svjour3}  
\smartqed  
\RequirePackage{graphicx}
\usepackage{amsmath}
\usepackage{url}
\journalname{}
\begin{document}

\title{A Quantitative Method for Evaluating Security Boundaries in Quantum Key Distribution Combined with Block Ciphers
}


\author{Xiaoming Chen\thanksref{eq,addr1,addr2}
        \and
        Haoze Chen\thanksref{eq,addr1,addr2}
        \and
        Fei Xu\thanksref{addr2}
        \and
        Meifeng Gao \thanksref{addr2}
        \and
        Jianguo Xie \thanksref{addr3,addr2}
        \and
        Cheng Ye \thanksref{addr1,addr2}
        \and
        An Hua \thanksref{addr1,addr2}
        \and
        Shichang Jiang \thanksref{addr2}
        \and
        Jiao Zhao \thanksref{addr1}
        \and
        Minghan Li \thanksref{addr1,addr2}
        \and
        Feilong Li \thanksref{addr1,addr2}
        \and
        Yajun Miao\thanksref{cor,addr1,addr2}
        \and
        Wei Qi\thanksref{cor,addr1,addr2}
}

\thankstext{eq}{These authors contributed equally to this work.}
\thankstext{cor}{Corresponding authors: Wei Qi (qiwei@qtict.com),
Yajun Miao (miaoyajun@qtict.com).}


\institute{CAS Quantum Network Co., Ltd., Shanghai, China. \label{addr1}
           \and
           Anhui CAS Quantum Network Co., Ltd., Hefei, China. \label{addr2}
           \and
           University of Science and Technology of China, Hefei, China.\label{addr3}
}


\maketitle

\begin{abstract}
With the rapid development of quantum computing, classical cryptography systems are increasingly vulnerable to security threats, thereby highlighting the urgency of constructing architectures that are resilient to quantum computing attacks. While Quantum Key Distribution (QKD) offers security with information-theoretic guarantees, its relatively low key generation rate necessitates integration with classical cryptographic techniques, particularly block ciphers such as AES and SM4, to facilitate practical applications. However, when a single QKD-key is employed to encrypt multiple data blocks, the reduction in cryptographic security strength has not yet been quantitatively analyzed. In this work, we focus on the security strength in the application scenario where QKD is combined with block ciphers. We propose a quantitative evaluation method for the security benefits of the QKD-key renewal period, aiming to provide a precise measure of the cryptographic security strength in such hybrid systems. Our method is based on concrete security paradigm of block cipher modes of operation. We demonstrate that under practical security level requirements, for files consisting of specific blocks, rekeying $k$ times can provide an additional $log_2k$ to $2log_2k$ bits of security. Our research offers a novel perspective on balancing the security and efficiency of QKD-based encryption.
\keywords{Quantum Key Distribution \and Block cipher \and Key reuse \and CTR mode \and CBC mode \and ECBC-MAC mode}
\end{abstract}

\section{Introduction}
\label{intro}
With the rapid development of quantum computing technology, classical cryptographic systems are facing increasingly realistic security threats, highlighting the urgency to upgrade current security architectures to resist attacks from quantum computers. Quantum key distribution (QKD) provides information-theoretic security and is widely regarded as a foundational technology for constructing information security assurance systems in the quantum era \cite{10.1145/382780.382781,PhysRevLett.85.441,10.1007/978-3-540-30576-7_21}. In theory, when combined with the One-Time Pad (OTP), QKD can provide unconditional security for data confidentiality. In practice, however, QKD-key generation rates typically remain on the order of Kbps, which is far below classical data transfer rates (Gbps–Tbps). Consequently, in practical deployments, QKD is integrated with conventional block ciphers  (such as AES \cite{10.1007/10721064_26,fips197} or SM4 \cite{rfc8998})  to protect bulk data under computational security. In such hybrid systems, the data security is strongly correlated with the QKD-key renewal period, yet this relationship has not been quantitatively characterized in existing literature. This motivates a systematic security analysis of QKD when combined with traditional block-cipher modules.

Bellare et al. \cite{bellare1997concrete} first analyzed various security notions for symmetric-key cryptography, such as Chosen-Plaintext Attack (CPA) and Chosen-Ciphertext Attack (CCA) security, within the concrete security framework. Based on the Birthday Attack \cite{von1939aufteilungs,Diffie1976new}, NIST SP 800-57 \cite{nist_sp80057_2020} provides qualitative suggestions or fixed quantitative thresholds for key management of cryptographic algorithms, such as key lifecycle and minimum security strength, but does not further specify the specific requirements for modes such as CTR and CBC.
Bellare et al. further demonstrated in \cite{bellare1997concrete} that the adversary advantage of block ciphers such as CTR and CBC under CPA can be attributed to the superposition of the distinguishing advantage between the underlying pseudo-random function (PRF) and a true random function, and the upper bound of the collision probability from birthday attack. From this, they derived the upper bound of the guessing advantage for these modes. Daniele et al. \cite{micciancio2018bit} proposed a new definition of bit security that is both general and quantitative. This definition covers both search and decision games, and establishes a tight reduction framework for cross-type primitives, providing a rigorous theoretical basis for bit security to become a cryptographic standard (such as key-length selection and security strength assessment). These results, along with the full-entropy property of QKD-keys, establish a solid foundation for analyzing the security boundaries of reusing a single QKD-key.

In this paper, we develop a computation model of QKD-key renewal period for multiple operating modes of block cipher algorithms. Given a specified mode and a target security strength, the QKD-key renewal period is derived based on the average security strength of the block cipher, the number of encrypted files, and the number of blocks per file. The resulting boundary is more precise than the rough $2^{n/2}$ estimate, thereby helping to avoid unnecessary key waste from overly frequent renewal while still meeting security requirements. Furthermore, we propose a quantitative evaluation method for the security benefit brought by QKD-key renewal. This method quantifies the security strength improvement (in bits) achieved by increasing the QKD-key renewal frequency under a given encryption mode. It also enables to quantitatively evaluate the security benefits of bringing quantum keys to encryption systems, which is conducive to establish the standards of QKD applications and large-scale promotion.
\section{Preliminaries}
\subsection{Quantum key distribution}
Quantum key distribution (QKD) leverages the principles of quantum mechanics to enable secure communication between two parties by negotiating a shared random key, which can be used for encryption and decryption. The field originated with the BB84 protocol proposed by Bennett and Brassard in 1984 \cite{BENNETT2014}.

The most crucial security feature of QKD is that any attempt by a third party to eavesdrop on the key information will be detected by the communicating parties. This security stems from the fundamental laws of quantum mechanics, specifically that an eavesdropper must perform a measurement on the quantum channel to acquire the key information, and this measurement introduces detectable anomalies. By using quantum superposition states to transmit key information, the communicating parties can verify signs of interference during the communication process and determine whether eavesdropping has occurred.

QKD is one of the important applications of quantum information science, providing a key distribution mechanism based on physical laws for modern cryptography. Traditional key transmission or key agreement schemes rely on assumptions based on mathematical problems such as integer factorization and discrete logarithms to establish security, however, the security of these schemes will be severely threatened if these computational problems are broken by more efficient algorithms or by adversaries with greater computational power (such as quantum computing adversaries). QKD can guarantee the security of the key even when facing an attacker with unlimited computing resources. Therefore, unlike traditional methods which rely on the assumptions of computational complexity, the security of QKD is rooted in the basic principles of quantum mechanics and can theoretically achieve information-theoretic security.

\subsection{CTR mode of operation}
\begin{figure}
    \includegraphics[width=0.5\textwidth]{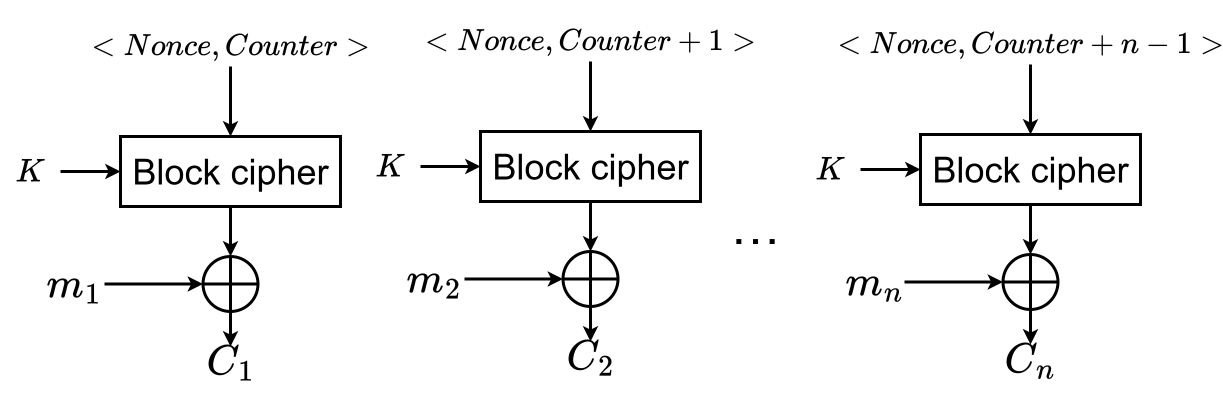}
\caption{CTR mode of operation}
\label{CTR}
\end{figure}

 Counter (CTR) mode \cite{diffie1979privacy} is a mode of operation that converts a block cipher algorithm into a stream cipher. It is characterized by high efficiency and parallelism in its application within block cryptography. As shown in Fig.\ref{CTR}, this mode uses a constantly changing counter as the input to the encryption algorithm, where the encrypted output of the counter is bitwise XORed with the plaintext block in each round to generate the ciphertext. After each encryption, the counter typically increments by a fixed value, providing a unique input for the next round.

 The primary advantages of CTR mode stem from its stream-cipher nature. It enables full parallelization in both encryption and decryption, as all keystream blocks can be generated independently, significantly enhancing throughput in high-performance applications. It also supports random access to ciphertext data, allowing any block to be decrypted without processing preceding blocks, which is particularly beneficial for encrypting storage devices or network packets. 

\subsection{CBC mode of operation}
\begin{figure}
    \includegraphics[width=0.5\textwidth]{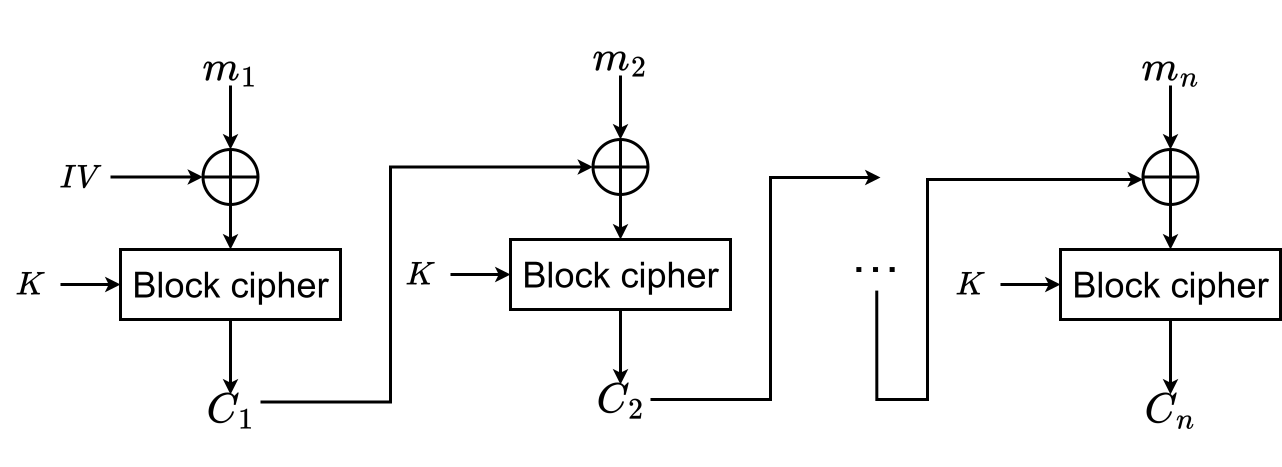}
\caption{CBC mode of operation}
\label{CBC}
\end{figure}

Cipher Block Chaining (CBC) mode \cite{nist1980fips81} is a widely used mode of operation in block cipher algorithms. As shown in Fig.\ref{CBC}, its core mechanism is to XOR the current plaintext block with the previous ciphertext block before inputting the result into the encryption algorithm to generate the current ciphertext block. At the start of the entire encryption process, a non-repeatable initialization vector (IV) is selected to ensure the randomness and security of the encrypted output.

CBC mode possesses strong diffusion capability, each ciphertext block depends not only on the current plaintext but also on all preceding ciphertext blocks, thereby enhancing its resistance to analysis. In the decryption phase, the recovery of plaintext depends on the current ciphertext and the previous ciphertext block. If an error occurs in a bit of the ciphertext, it will cause the corresponding plaintext block to be completely incorrect, and the corresponding bit of the subsequent plaintext block may also be affected, but it will not affect blocks further down the chain, demonstrating a certain degree of error localization property.
\subsection{ECBC-MAC mode of operation}
 Encrypted Cipher Block Chaining Message Authentication Code (ECBC-MAC) is introduced as an enhancement to address the security vulnerabilities of standard CBC-MAC when handling variable-length messages. CBC-MAC is a widely adopted Message Authentication Code construction based on the CBC mode, primarily used in symmetric-key cryptography for data integrity protection. In this mode, the entire message is segmented into fixed-length blocks, processed sequentially using CBC encryption, and only the final ciphertext block is retained as the authentication tag output. This tag is then used to verify that the data has not been tampered with during transmission \cite{bellare2000cbc}.

\begin{figure}    \includegraphics[width=0.5\textwidth]{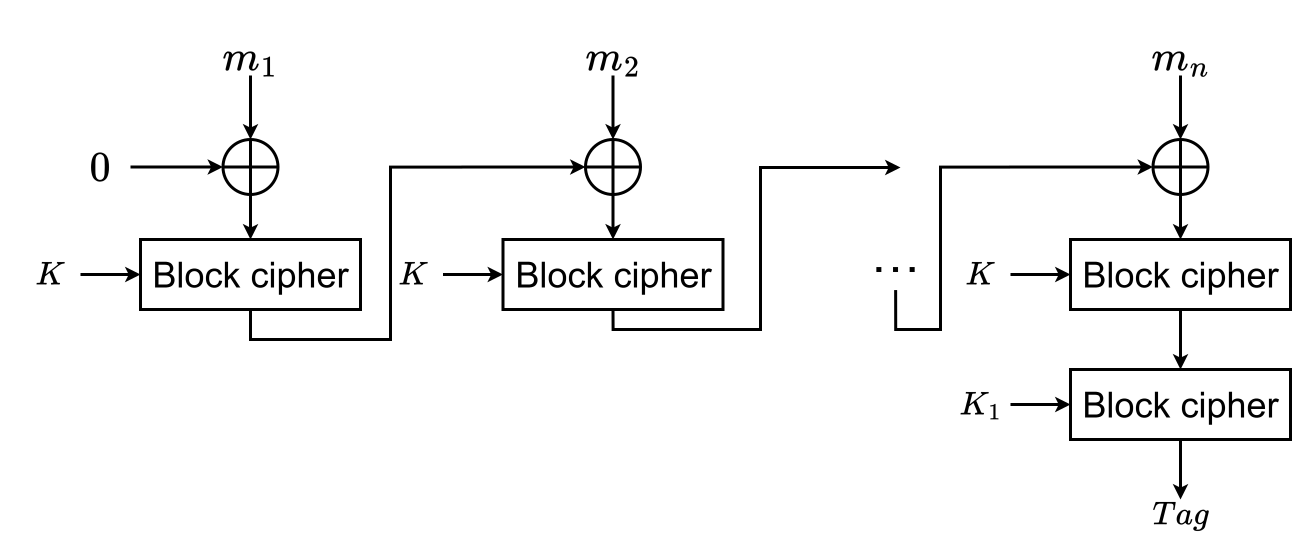}
\caption{ECBC-MAC mode of operation}
\label{ECBC-MAC}
\end{figure}

Although CBC-MAC is proven secure for fixed-length messages, it suffers from the Length Extension Attack vulnerability when applied to variable-length messages. ECBC-MAC resolves this flaw by introducing an additional encryption step at the end of the process. As shown in Fig.\ref{ECBC-MAC}, the fundamental approach is to first process the message in CBC mode using a key $K$  to obtain an intermediate value $t=F_{CBC}$, and another key $K_1$ is employed to encrypt $t$, ultimately generating $tag=E(K_1,t)$. Both $F$ and $E$ can use the same secure PRF (or block cipher algorithm).

Furthermore, to securely manage situations where the message length is not an exact multiple of the block size (i.e., padding), researchers have proposed a variety of enhanced structures, such as the three-key XCBC-MAC \cite{bellare2000cbc} and the CMAC mode \cite{dworkin2003cmac} recommended by NIST, in order to improve efficiency and simplify key management.

\subsection{Asymptotic security}
The security of an encryption scheme is typically described by the probability of an adversary successfully compromising the system under a specific attack model. Common attack types include the Chosen-Plaintext Attack (CPA) and the Chosen-Ciphertext Attack (CCA). The CPA model adopted in this paper means that the adversary, without knowing the key, can access the encryption algorithm's query interface and adaptively submit multiple plaintexts to obtain their corresponding ciphertexts. If the adversary makes at most $q$ encryption queries and the total attack time is at most $t$, the probability of successfully breaking the system can be expressed as $\epsilon(\lambda, t,q)$, where $\lambda$ is the security parameter. In most cases, there is a tight positive correlation between $q$ and $t$ (since the main part of $t$ is $q$), so these two parameters are often simplified into one.

Asymptotic Security refers to the following: For any natural number $c$, there exists a critical security parameter $\lambda_0$ such that when $\lambda > \lambda_0$ and the adversary's computational resources $t$ satisfy $t \leq \mathrm{poly}(\lambda)$ for a given $\lambda$, we have
$$
\epsilon(\lambda, t) < \frac{1}{\lambda^c}.
$$
This definition embodies the typical complexity assumption in theoretical cryptography—as long as a sufficiently large security parameter is selected and the adversary is restricted to a polynomial-time algorithm, the encryption scheme can achieve target security level. Within this framework, security is bounded by a negligible function of $\lambda$ \cite{goldwasser1984probabilistic,katz2014introduction}.

However, a key limitation of this method is that it only provides an "existence" guarantee for the aforementioned bound: although for any $c$, there exists some $\lambda_0$ such that the inequality holds, the model does not provide a concrete correspondence between $\lambda_0$ and the parameters $c$ or the attack resource $t$. Furthermore, it is impossible to infer the security parameters that should be used in a practical system from the expected upper limit of $\epsilon$. Therefore, in practical deployment and system design, asymptotic security often struggles to provide direct guidance for parameter selection \cite{bellare1997concrete,rogaway2004practice}.
\subsection{Concrete security}
Unlike asymptotic security, the concrete security approach is not satisfied with considering the limiting situation of $\epsilon(\lambda,t)$; instead, it directly provides an explicit expression for the adversary's success probability $\epsilon(\lambda, t)$ given specific resource constraints. This model was first systematically proposed by Bellare et al. in their security analysis of symmetric encryption \cite{bellare1997concrete} and is widely used for the security evaluation of practical cryptographic protocols.

In concrete security analysis, the maximum resource limit $t_{\mathrm{max}}$ for the adversary and the maximum tolerable success probability $\epsilon_{\mathrm{max}}$ are usually preset. Then, under the constraint that
$$
\epsilon(\lambda, t_{\mathrm{max}}) \leq \epsilon_{\mathrm{max}}
$$
the smallest security parameter $\lambda$ that satisfies the condition is determined,  which is the required parameter length for system deployment.

Concrete security allows researchers to weigh security and cost based on the actual threat model faced by the system and provides a basis for parameter selection in practical system deployment, thus being widely adopted by standardization organizations (such as NIST). Furthermore, concrete security proofs also support the quantification of "security reductions," for example, by connecting an attacker's advantage over a cryptographic primitive to their success probability on an underlying hard problem (such as the discrete logarithm or RSA inversion), resulting in a clear security loss factor (reduction loss).

In comparison, while asymptotic security possesses theoretical clarity, it fails to quantify the relationship between the security parameter and the actual attack cost. The concrete security model, conversely, is better suited for the analysis and evaluation of cryptographic systems in real-world engineering \cite{katz2014introduction,rogaway2004practice}.

\section{Security benefit quantitative evaluation method}
Block ciphers include several modes of operation, such as CTR, CBC, and ECBC-MAC. When calculating the problem of key reuse in block ciphers, information-\\theoretically secure keys are computationally feasible because their security strength and bit length are equivalent. Keys generated by a QKD network rely on the laws of physics and possess information-theoretic security properties. Aiming at the integrated use of QKD-keys generated by QKD networks with various block cipher modes of operation, this paper proposes a quantitative evaluation method for the QKD-key renewal period and security benefits. The three modes that primarily use this method are introduced here: CTR encryption mode, CBC encryption mode, and ECBC-MAC mode.
\subsection{Symbols}
\begin{table*}
     \caption{List of symbols}
     \centering
\begin{tabular}{ll}
\hline
\noalign{\smallskip}
Symbol & Descriptions \\
\noalign{\smallskip}
\hline\noalign{\smallskip}
$\mathcal{E}=(E,D)$ & Includes encryption algorithm $E$ and decryption algorithm $D$ \\
$\lambda$ & Security parameter \\
$t$ & Adversary's attack time cost \\
$q$ & Maximum number of times adversary $\mathcal{A}$ can query the encryption system \\
$\mathcal{B}^\ast$ & The current optimal attack algorithm \\
$\mathrm{CPAadv}[\mathcal{A,E}]$ & Adversary $\mathcal{A}$'s distinguishing advantage against $\mathcal{E}$ \\
$\epsilon(\lambda,t)$ & Adversary $\mathcal{A}$'s guessing advantage against $\mathcal{E}$ \\
$\rho(\lambda,t)$ & Distinguishing advantage against a PRF \\
$Q^\ast$ & Number of files encrypted by the key, i.e., the QKD-key renewal period \\
$C_{\mathrm{QKD\text{-}key}}$& The cost of a QKD-key\\
\noalign{\smallskip}
\hline
\end{tabular}
    \label{List of Symbols}
\end{table*}
Table \ref{List of Symbols} lists the symbols in this paper.  $\mathcal{E}=(E,D)$ is the given block cipher algorithm, which includes the encryption algorithm $E$ and the decryption algorithm $D$, and $D(k,E(k,m))=m$. $\lambda$ is the security parameter, representing the system's security level; theoretically, a larger security parameter results in higher system security, but efficiency usually decreases. $t$ is the maximum computational time that an adversary spends to break the cryptographic system. $q$ is the maximum number of time that the adversary can interact with the encryption system. $\mathcal{B}^\ast$ is the most efficient known attack algorithm against the block cipher algorithm $\mathcal{E}$. $\mathrm{CPAadv}[\mathcal{A,E}]$ is a probability value ranging from $[0, 1]$, used to precisely measure the adversary's capability to break the encryption scheme $\mathcal{E}$ under the CPA model. $\epsilon(\lambda,t)$ describes the maximum attack advantage that can be obtained by all adversaries running within time $t$; this advantage is a function of the security parameter $\lambda$ and the attack time $t$. $\rho(\lambda,t)$ specifically refers to the security of the  PRF, representing the maximum advantage in distinguishing a PRF from a completely random function for any adversary within time t. $Q^\ast$ refers to the maximum number of files or messages encrypted by the same key before a new encryption key is renewed, and it can also be used to represent the QKD-key renewal period. $C_{\mathrm{QKD\text{-}key}}$ refers to the cost of a QKD-key.
\subsection{Security level}
Given the block cipher algorithm $\mathcal{E}=(E,D)$ and its security definition under CPA, the distinguishing advantage function of the adversary (distinguisher algorithm) $\mathcal{A}$ is denoted as $\mathrm{CPAadv}[\mathcal{A,E}]$. Its guessing advantage is defined as\cite{bellare1997concrete}:
\begin{equation}\label{Formula1}
\epsilon(\lambda,t)=\frac{1}{2}\cdot \mathrm{CPAadv}[\mathcal{A,E}]
\end{equation}
The security strength of the encryption system is quantified using the following metrics:
\begin{itemize}
    \item Worst-Case Security Strength\cite{lee2022bit}:
    \begin{equation}\label{Formula2}
     \max_{t \leq q} \left\{ \log_2 \frac{1}{\epsilon(\lambda, t)} \right\}\ (\mathrm{bit})
    \end{equation}
    This value characterizes the system's resistance to the optimal attack strategy.
    \item Average-Case Security Strength\cite{micciancio2018bit}:
    \begin{equation}\label{Formula3}
     \min_{t \leq q} \left\{ \log_2 \frac{t}{\epsilon(\lambda, t)} \right\}\ (\mathrm{bit})
    \end{equation}
    This value reflects the system's robustness under typical attack scenarios.
\end{itemize}    
\subsection{Quantitative model of QKD-key renewal period and security benefits in CTR mode.}
\begin{figure}[h]
\begin{center}
    \includegraphics[width=0.5\textwidth]{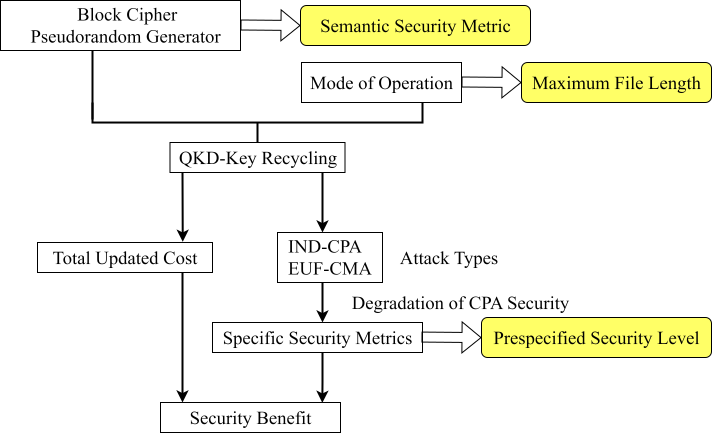}
\end{center}
\caption{Calculation flowchart}
\label{process}
\end{figure}
As shown in Fig.\ref{process} , the model includes two steps: parameter selection and metric calculation. This means that one must first determine the distinguishing advantage of the cryptographic module and the maximum number of blocks per single encrypted file. Then, after determining the security strength parameters of the encryption system, the metric calculation is performed.

\noindent\textbf{Step 1: parameter selection}\\
\noindent (1) Distinguishing advantage of the cryptographic module $F$:

The cryptographic module $F$ (which typically employs a standardized block cipher algorithm such as AES or SM4, or derivative schemes built on them) serves as the basic component. Its concrete security metrics have been validated through long-term cryptanalysis, reaching an agreement in academia. If the distinguishing advantage of the current internationally recognized optimal attack algorithm $\mathcal{B}^\ast$ is denoted as $\mathrm{PRFadv}[\mathcal{B}^\ast,F]$, the guessing advantage is half of the distinguishing advantage, i.e.,
\begin{equation}\label{Formula4}
\rho(\lambda,t)=\frac{1}{2}\mathrm{PRFadv}[\mathcal{B}^\ast,F]
\end{equation}
The concrete security of the PRF is determined based on the advantage of the guessing attack.

The value of $\rho(\lambda,t)$ is essentially fixed when the block cipher algorithm is solicited. As the analysis of $F$ in the cryptography community continues to deepen (reflected in the efficiency improvement of the attack algorithm $\mathcal{B}^\ast$), $\rho(\lambda,t)$ exhibits a strictly monotonically increasing trend. For widely deployed core algorithms, their average security strength has a clear infimum $$s_{\mathrm{min}}\approx\mathrm{inf}_{t>0}\{\log_{2}\frac{t}{\rho(\lambda,t)}\}$$ (bits). This value characterizes the average computational cost required for the attack algorithm to succeed (i.e., the average security strength), and it provides the theoretical basis for the engineering calculation of concrete security boundaries.

\noindent (2) Maximum number of blocks per single file, $l$:

This parameter is determined according to the actual application scenario and defines the maximum number of encrypted blocks that a single file can be divided into. In the subsequent security benefit evaluation, to simplify the analysis (for standardization needs), it is often uniformly assumed that all files reach the maximum block length $l$.

\noindent (3) Security level of the encryption scheme, $\epsilon_{\mathrm{max}}$:

This is determined by the security requirements of the application system and must be confirmed by consulting between the cryptographic system designer and the user.

\noindent\textbf{Step 2: metric calculation}\\
\noindent (1) Evaluation method for QKD-key renewal period $Q^\ast$ (maximum number of files encrypted by a single QKD-key):

\noindent By solving the following integer programming problem \cite{bellare1997concrete}:
\begin{equation}\label{Formula5}
\mathrm{max}\ Q:\ \mathrm{s.t.\ \rho (\lambda,Ql)+\frac{2Q^2l}{N}\leq\epsilon_{\mathrm{max}}}
\end{equation}
\noindent where $N$ is the size (number of elements) of the domain space of $F$, typically satisfying $N=2^\lambda$, and $Q$ is the number of plaintexts chosen by the attacker (queries to the encrypting machine).

\noindent Since the distinguishing advantage $\rho(\lambda,Ql)$ is monotonically increasing with $Q$ (the success rate of the attack increases as the query volume grows), this is equivalent to solving the equation:
\begin{equation}\label{Formula6}
\rho (\lambda,Ql)+\frac{2Q^2l}{N}=\epsilon_{\mathrm{max}}
\end{equation}
\noindent For mainstream block cipher algorithms (such as SM4, AES, 3DES) or the PRFs constructed based on them, the infimum of the average security strength $s_{\mathrm{min}}$ is known, approximates a constant within a certain range, and satisfies $s_\mathrm{min}\sim O(N)$. At this point:
\begin{equation}
\frac{Ql}{\rho(\lambda,Ql)}\approx s_{\mathrm{min}} \quad
\end{equation}
It can be deduced that:
\begin{equation}\label{Formula7}
\rho(\lambda,Ql)\approx \frac{Ql}{s_{\mathrm{min}}}
\end{equation}
\noindent Combining Equation (\ref{Formula6}) and Equation (\ref{Formula7}), we obtain the simplified quadratic equation for $Q$:
\begin{equation}\label{Formula8}
\frac{Ql}{s_{\mathrm{min}}}+\frac{2Q^2l}{N}=\epsilon_{\mathrm{max}}
\end{equation}
\noindent By solving Equation (\ref{Formula8}), we can get the positive integer root $Q^\ast$.

$Q^\ast$ represents the theoretical maximum number of files that a single QKD-key can encrypt, ensuring the preset security strength that the CTR mode encryption system can achieve.

\noindent (2) Quantitative improvements in security strength by QKD-key renewal:

\noindent Based on the result of the QKD-key renewal period $Q^\ast$, the worst-case security strength of the encryption system is:
\begin{equation}\label{Formula9}
-\log_{2}(\frac{Q^\ast l}{s_{\mathrm{min}}}+\frac{2(Q^\ast)^2l}{N})=-\log_{2}\epsilon_{\mathrm{max}}
\end{equation}


According to Equation (\ref{Formula9}), if QKD-key renewal is performed uniformly $k$ times (i.e., the number of files encrypted by a single QKD-key is reduced to $Q^\ast/k$), the security strength improvement is calculated as:
\begin{equation}\label{Formula11}
\begin{aligned}
&-\log_{2}\!\Biggl(
\frac{(Q^\ast/k) l}{s_{\mathrm{min}}}
+
\frac{
2 (Q^\ast/k)^{2} l
}{
N
}
\Biggr) \\
&\quad
-\Biggl[
-\log_{2}\!\Biggl(
\frac{Q^\ast l}{s_{\mathrm{min}}}
+
\frac{
2 {Q^\ast}^{2} l
}{
N
}
\Biggr)
\Biggr] \\
&=
\log_{2} k
+
\log_{2}\!\Biggl(
1+
\frac{
2 (k-1) Q^\ast \cdot s_{\mathrm{min}}
}{
k N
+
2 Q^\ast \cdot s_{\mathrm{min}}
}
\Biggr)
\;(\text{bit})
\end{aligned}
\end{equation}

\noindent \textbf{Conclusion: Uniformly executing $k$ QKD-key renewals can improve the security strength by  $\log_{2}k$ to $2\log_{2}k$ bits (strictly greater than 
$\log_{2}k$ bits and less than $2\log_{2}k$ bits).}

\noindent (3) Quantitative evaluation method for security benefits of QKD-key reuse:

For an encryption system using a highly secure PRF $F$ (or block cipher algorithm) with the CTR mode, when the maximum number of files encrypted by a single QKD-key is $Q^\ast$, the basic security strength guarantee is $\log_{2}\frac{1}{\epsilon_\mathrm{max}}$ (bits). By introducing ideal keys(QKD-keys) and performing $k$ uniform QKD-key renewals, for a moderate value of $k$, an approximate security strength improvement of
\begin{equation}\label{Formula12}
\log_{2}k+\log_{2}(1+\frac{2(k-1)Q^\ast \cdot s_{\mathrm{min}}}{kN+2Q^\ast \cdot s_{\mathrm{min}}})
\end{equation}
\noindent bits can be achieved.

From this, the computation model for the relationship between key cost and security strength improvement through key reuse (with k uniform QKD-key renewals) can be represented by "Security benefits", which equals the following value:
\begin{equation}\label{Formula13}
\frac{Q^\ast\times [\log_{2}k+\log_{2}(1+\frac{2(k-1)Q^\ast \cdot s_{\mathrm{min}}}{kN+2Q^\ast \cdot s_{\mathrm{min}}})]}{k\times C_{\mathrm{QKD\text{-}key}}};
\end{equation}
where the cost of a QKD-key is determined by the service price of the Quantum Key Distribution infrastructure.
\subsection{Quantitative model of QKD-key renewal period and security benefits in CBC mode.}
Given that the security conclusions and provable security framework of the CBC mode are highly similar to those of the CTR mode, the security benefit evaluation for QKD-key can follow the CTR mode evaluation framework, requiring adjustment only of a few core parameters. This section will briefly describe the evaluation method under the CBC mode, primarily clarifying the key differences from the CTR mode.

\noindent\textbf{Step 1: parameter selection}

\noindent (1) Distinguishing advantage of the cryptographic module (block cipher algorithm) $E$:

\noindent The value of $\rho(\lambda,t)$ is essentially fixed when the block cipher algorithm is solicited. For widely deployed core algorithms, their average security strength has a clear infimum $s_{\mathrm{min}}\approx\mathrm{inf}_{t>0}\{\log_{2}\frac{t}{\rho(\lambda,t)}\}$ (bits). This value characterizes the minimum resource consumption required for an attacker to succeed, representing the average security strength of the algorithm against attacks.

\noindent (2) Maximum number of blocks per single file, $l$:

\noindent This parameter is set according to the actual application scenario and defines the upper limit on the number of blocks when a single file is divided into encrypted blocks. In the subsequent security benefit evaluation, to simplify the analysis process, it is assumed that all files reach the maximum block length $l$.

\noindent (3) Security level of the encryption scheme, $\epsilon_{\mathrm{max}}$:

\noindent This is determined by the security requirements of the application system and is generally confirmed through consultation between the cryptographic system designer and the user.

\noindent\textbf{Step 2: metric calculation}

\noindent (1) Quantitative improvements in security strength by QKD-key renewal:

The security evaluation method is almost the same as the evaluation approach for the CTR mode. The quadratic equation for $Q$ is solved as follows:
\begin{equation}\label{Formula14}
\frac{Ql}{s_{\mathrm{min}}}+\frac{2Q^2l^2}{N}=\epsilon_{\mathrm{max}}
\end{equation}
\noindent The positive integer root $Q^\ast$ is obtained, which represents the theoretical maximum number of files that a single QKD-key can encrypt, ensuring the preset security strength that the CBC mode encryption system can achieve.

\noindent Based on the result of the QKD-key renewal period $Q^\ast$, the worst-case security strength of the encryption system is:
\begin{equation}\label{Formula15}
-\log_{2}(\frac{Q^\ast l}{s_{\mathrm{min}}}+\frac{2(Q^\ast)^2l^2}{N})=-\log_{2}\epsilon_{\mathrm{max}}
\end{equation}


According to Equation (\ref{Formula15}), if QKD-key renewal is performed uniformly $k$ times (i.e., the number of files encrypted by a single QKD-key is reduced to $Q^\ast/k$), the security strength improvement is calculated as:
\begin{equation}\label{Formula17}
\begin{aligned}
&-\log_{2}\!\Biggl(
\frac{(Q^\ast/k) l}{s_{\mathrm{min}}}
+
\frac{
2 (Q^\ast/k)^{2} l^{2}
}{
N
}
\Biggr) \\
&\quad
-\Biggl[
-\log_{2}\!\Biggl(
\frac{Q^\ast l}{s_{\mathrm{min}}}
+
\frac{
2 {Q^\ast}^{2} l^{2}
}{
N
}
\Biggr)
\Biggr] \\
&=
\log_{2} k
+
\log_{2}\!\Biggl(
1+
\frac{
2 (k-1) Q^\ast l \cdot s_{\mathrm{min}}
}{
kN
+
2 Q^\ast l \cdot s_{\mathrm{min}}
}
\Biggr)
\;(\text{bit})
\end{aligned}
\end{equation}

\noindent \textbf{Conclusion: Uniformly executing $k$ QKD-key renewals can improve the security strength by  $\log_{2}k$ to $2\log_{2}k$ bits (strictly greater than 
$\log_{2}k$ bits and less than $2\log_{2}k$ bits).}

\noindent (2) Quantitative evaluation method for security benefits of QKD-key reuse:

 For an encryption system using a highly secure block cipher algorithm in CBC mode, when the maximum number of files encrypted by a single QKD-key is $Q^\ast$, the basic security strength is $\log_{2}\frac{1}{\epsilon_\mathrm{max}}$ (bits). By introducing ideal QKD-keys and performing $k$ uniform QKD-key renewals, for a moderate value of $k$, an approximate security strength improvement of
\begin{equation}\label{Formula18}
\log_{2}k+\log_{2}(1+\frac{2(k-1)Q^\ast l \cdot s_{\mathrm{min}}}{kN+2Q^\ast l \cdot s_{\mathrm{min}}})
\end{equation}
\noindent bits can be achieved.

\noindent The computation model for the relationship between key cost and security strength improvement through key reuse (with k uniform QKD-key renewals) can be represented by "Security benefits", which equals the following value:
\begin{equation}\label{Formula19}
\frac{Q^\ast\times [\log_{2}k+\log_{2}(1+\frac{2(k-1)Q^\ast l \cdot s_{\mathrm{min}}}{kN+2Q^\ast l \cdot s_{\mathrm{min}}})]}{k\times C_{\mathrm{QKD\text{-}key}}};
\end{equation}
where the cost of a single QKD-key is determined through consultation between the operator of the quantum cryptography infrastructure and the user.

\subsection{Quantitative model of QKD-key renewal period and security benefits in ECBC-MAC mode.}
Given that the security conclusions and provable security framework of the ECBC-MAC mode are highly similar to those of the CBC mode, the security benefit evaluation for QKD-keys follows the CBC mode evaluation framework, requiring adjustment only of a few core parameters. This section will briefly describe the evaluation method under the ECBC-MAC mode, primarily clarifying the key differences from the CBC mode.

\noindent\textbf{Step 1: parameter selection}

\noindent (1) Distinguishing advantage of the cryptographic module $F$:

The value of $\rho(\lambda,t)$ is essentially fixed when the block cipher algorithm is solicited. For widely deployed core algorithms, their average security strength has a clear infimum $s_{\mathrm{min}}\approx\mathrm{inf}_{t>0}\{\log_{2}\frac{t}{\rho(\lambda,t)}\}$ (bits). This value characterizes the minimum security strength of the algorithm against unit attack cost and provides the theoretical basis for the engineering deployment of concrete security boundaries.

\noindent (2) Maximum number of blocks per single file, $l$:

This parameter is set according to the actual application scenario and defines the upper limit on the number of encrypted blocks into which a single file is divided. In the subsequent security benefit evaluation, to simplify the analysis (for standardization needs), it is often uniformly assumed that all files reach the maximum block length $l$.

\noindent (3) Security level of the encryption scheme, $\epsilon_{\mathrm{max}}$:

This is determined by the security requirements of the application system and must be confirmed by consulting between the cryptographic system designer and the user.

\noindent\textbf{Step 2: metric calculation}

\noindent (1) Quantitative improvements in security strength by QKD-key renewal:

\noindent The security evaluation method is fundamentally the same as the evaluation approach for the CBC mode. The quadratic equation for $Q$ is solved as follows:

\noindent 1. Two keys $k_1, k_2$ are used.

\noindent 2. The quadratic equation is:
\begin{equation}\label{Formula20}
2\frac{Ql}{s_{\mathrm{min}}}+\frac{Q^2l^2+Q^2+2}{2N}=\epsilon_{\mathrm{max}}
\end{equation}
\noindent The positive integer root $Q^\ast$ is obtained, which represents the theoretical maximum number of files that a single QKD-key can encrypt, ensuring the preset security strength that the ECBC-MAC mode encryption system can achieve.

Based on the result of the QKD-key renewal period $Q^\ast$, the worst-case security strength of the encryption system is:
\begin{equation}\label{Formula21}
-\log_{2}(2\frac{Ql}{s_{\mathrm{min}}}+\frac{Q^2l^2+Q^2+2}{2N})=-\log_{2}\epsilon_{\mathrm{max}}
\end{equation}



According to Equation (\ref{Formula21}), if QKD-key renewal is performed uniformly $k$ times (i.e., the number of files encrypted by a single QKD-key is reduced to $Q^\ast/k$), the security strength improvement is calculated as:
\begin{equation}\label{Formula23}
\begin{aligned}
&-\log_{2}\!\Biggl(
2 \frac{(Q^\ast/k) l}{s_{\mathrm{min}}}
+
\frac{
(Q^\ast/k)^{2} l^{2}
+
(Q^\ast/k)^{2}
+
2
}{
2N
}
\Biggr) \\
&\quad
-\Biggl[
-\log_{2}\!\Biggl(
2 \frac{Q^\ast l}{s_{\mathrm{min}}}
+
\frac{
{Q^\ast}^{2} l^{2}
+
{Q^\ast}^{2}
+
2
}{
2N
}
\Biggr)
\Biggr] \\
&=
\log_{2} k
+
\log_{2}\!\Biggl(
1+
\frac{
{Q^\ast}^{2}(l^{2}+1)\!\left(1-\tfrac{1}{k}\right)
+2(1-k)
}{
4N \tfrac{Q^\ast l}{s_{\mathrm{min}}}
+\tfrac{{Q^\ast}^{2}(l^{2}+1)}{k}
+2k
}
\Biggr)
\end{aligned}
\end{equation}

\noindent \textbf{Conclusion: Uniformly executing $k$ QKD-key renewals can improve the security strength by  $\log_{2}k$ to $2\log_{2}k$ bits (strictly greater than 
$\log_{2}k$ bits and less than $2\log_{2}k$ bits).}

\noindent (2) Quantitative evaluation method for security benefits of QKD-key reuse:

For an encryption system using a highly secure PRF $F$ (or block cipher algorithm) with the ECBC-MAC mode, when the maximum number of files encrypted by a single QKD-key is $Q^\ast$, the basic security strength guarantee is $\log_{2}\frac{1}{\epsilon_\mathrm{max}}$ (unit: bits). By introducing ideal (QKD) keys and performing $k$ uniform QKD-key renewals, for a moderate value of $k$, an approximate security strength improvement of
\begin{equation}\label{Formula24}
\log_{2}k+\log_{2}(1+\frac{{Q^\ast}^2(l^2+1)(1-\frac{1}{k})+2(1-k)}{4N\frac{Q^\ast l}{s_{\mathrm{min}}}+\frac{{Q^\ast}^2(l^2+1)}{k}+2k})
\end{equation}
\noindent bits can be achieved.

The computation model for the relationship between key cost and security strength improvement through key reuse (with k uniform QKD-key renewals) can be represented by "Security benefits", which equals the following value:
\begin{equation}\label{Formula25}
\begin{aligned}
\quad
\frac{
Q^{\ast}
\Bigl[
\log_{2} k
+
\log_{2}\!\Bigl(
1+
\frac{
{Q^\ast}^{2}(l^{2}+1)\!\left(1-\tfrac{1}{k}\right)
+2(1-k)
}{
4N \tfrac{Q^\ast l}{s_{\mathrm{min}}}
+\tfrac{{Q^\ast}^{2}(l^{2}+1)}{k}
+2k
}
\Bigr)
\Bigr]
}{
k \times C_{\mathrm{QKD\text{-}key}}
}
\end{aligned}
\end{equation}
where the cost of a single QKD-key is determined through consultation  between the operator of the quantum cryptography infrastructure and the user.

\section{Calculation and analysis of specific examples}

Based on the security benefit evaluation method described above, we use the CTR mode of the AES-128 algorithm combined with QKD-keys to analyze the specific cost required for security improvement. $\mathcal{E}=(E,D)$ is the AES block cipher algorithm, with both the block length and key length being 128 bits. Here, $\mathcal{E}^\prime=(E^\prime,D^\prime)$ is the CTR mode of the AES block cipher algorithm, $\mathcal{E}^{\prime\prime}=(E^{\prime\prime},D^{\prime\prime})$ is the CBC mode of the AES block cipher algorithm, and $\mathcal{E}^{\prime\prime\prime}=(E^{\prime\prime\prime},D^{\prime\prime\prime})$ is the ECBC-MAC mode of the AES block cipher algorithm. $l$ is the number of blocks, and $Q$ is the number of encrypted files.

\noindent\textbf{Step 1: parameter assignment}

\noindent (1) The security parameter $\lambda$ of the AES block cipher is 128. We assume its average security strength is 126 bits\cite{AES126}, i.e., $s_{\mathrm{min}}\geq 2^{126}$.

\noindent (2) Assume the maximum length of a single file to be encrypted is 1.5 KB, thus the number of blocks $l=(2^{10}\times 2^3 \times 1.5)/2^7=96$.

\noindent (3) Ensure that $\mathcal{E}^\prime, \mathcal{E}^{\prime\prime}, \mathcal{E}^{\prime\prime\prime}$ have a security strength of 80 bits, i.e., $\epsilon_{\mathrm{max}}=1/2^{80}$.

\noindent\textbf{Step 2: metric calculation}

\noindent\textbf{CTR mode calculation}

\begin{figure*}[htbp]
\centering
\includegraphics[width=0.8\textwidth]{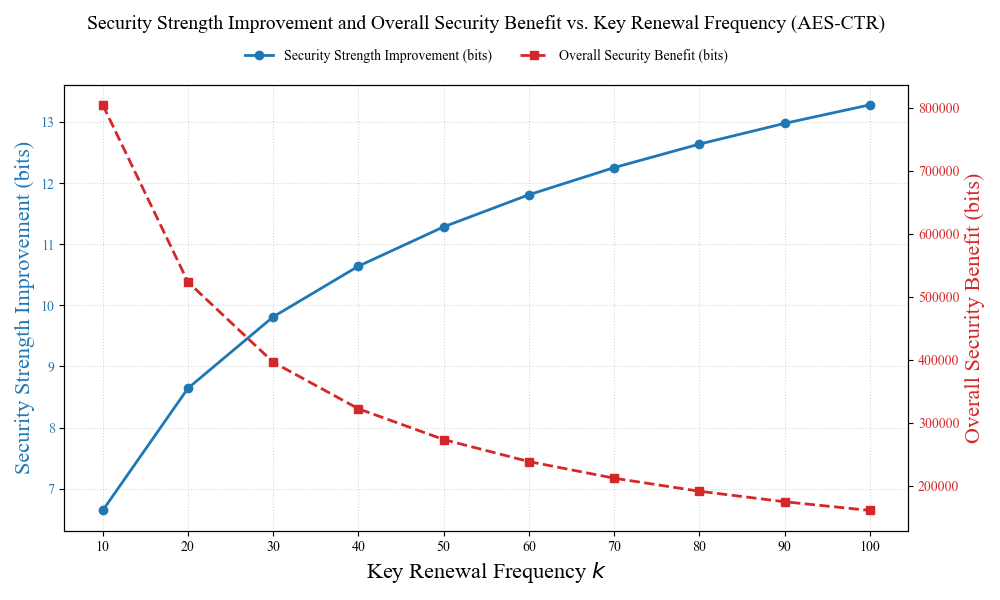} 
\caption{Relationship between security improvement and QKD-key renewal period in CTR mode}
\label{CTR-key}
\end{figure*}

\noindent (1) Calculate the maximum QKD-key renewal period $Q^\ast$, by solving the inequality:
\begin{equation}\label{Formula26}
\frac{96\times Q^\ast}{2^{126}}+\frac{2\times 96\times (Q^\ast)^2}{2^{128}}\leq \frac{1}{2^{80}}
\end{equation}
\noindent We obtain $Q^\ast=1210790$. This means each QKD-key can encrypt up to $1210790$ files, and the maximum encrypted data volume is $1210790\times 1.5\mathrm{KB}=1816185\mathrm{KB}\\ \approx 1774\mathrm{MB}$. Therefore, the key should be renewed after each QKD-key encrypts 1774 MB of data.

\noindent (2) Assume $k=2$, meaning $2$ uniform QKD-key renewals are executed (i.e., the number of files encrypted by a single QKD-key is reduced to $Q^\ast/2$). The security strength improvement is:
\begin{equation}\label{Formula27}
1+\log_{2}(1+\frac{Q^\ast\cdot s_{\mathrm{min}}}{N+Q^\ast\cdot s_{\mathrm{min}}})\approx 1.999923
\end{equation}

\noindent (3) By estimating the cost of replacing a 128-bit QKD-key, we can calculate the security benefits gained from using QKD-keys:
\begin{equation}\label{Formula28}
\frac{Q^\ast\times [1+\log_{2}(1+\frac{Q^\ast\cdot s_{\mathrm{min}}}{N+Q^\ast\cdot s_{\mathrm{min}}})]}{2\times C_{\mathrm{QKD\text{-}key}}}=\frac{1210790\times 1.999923}{2\times C_{\mathrm{QKD\text{-}key}}}
\end{equation}

\noindent As shown in Fig.\ref{CTR-key},  assuming the cost of a single QKD-key is $1$, in CTR mode, increasing the frequency of rekeying (i.e., shortening the QKD-key renewal period) leads to an increase in security strength but a decrease in the overall security benefits.

\noindent\textbf{CBC mode calculation}

\begin{figure*}[htbp]
\centering
\includegraphics[width=0.8\textwidth]{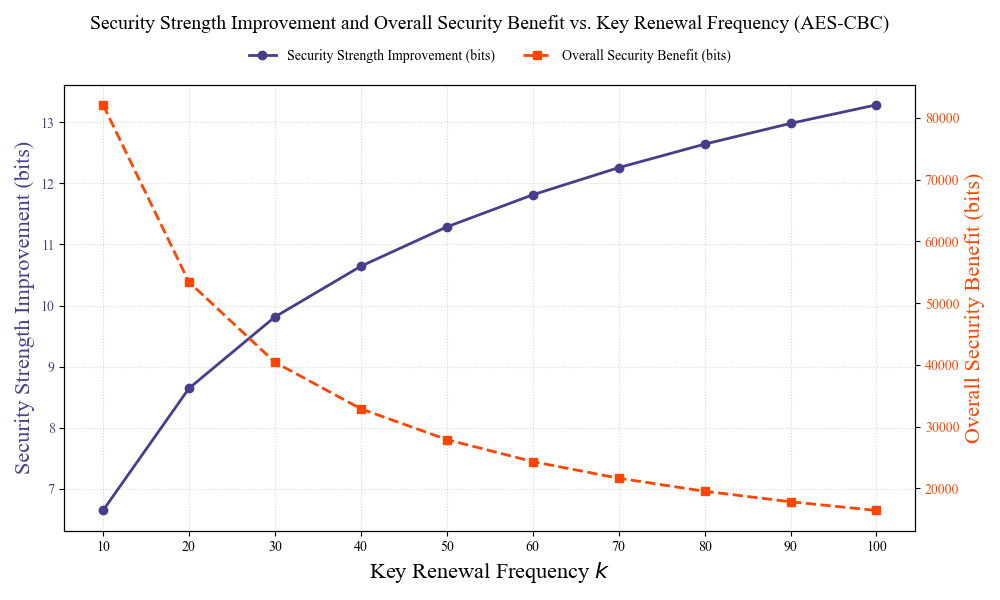} 
\caption{Relationship between security improvement and QKD-key renewal period in CBC mode}
\label{CBC-key}
\end{figure*}

\noindent (1) Calculate the maximum QKD-key renewal period $Q^\ast$, by solving the inequality:
\begin{equation}\label{Formula29}
\frac{96\times Q^\ast}{2^{126}}+\frac{2\times 96^2\times (Q^\ast)^2}{2^{128}}\leq \frac{1}{2^{80}}
\end{equation}
\noindent We obtain $Q^\ast=123577$. The maximum encrypted data volume is $123577\times 1.5\mathrm{KB}=185365\mathrm{KB}\approx 181\mathrm{MB}$. Therefore, the key should be renewed after each QKD-key encrypts 181 MB of data.

\noindent (2) Assume $k=2$, meaning $2$ uniform QKD-key renewals are executed (i.e., the number of files encrypted by a single QKD-key is reduced to $Q^\ast/2$). The security strength improvement is:
\begin{equation}\label{Formula30}
1+\log_{2}(1+\frac{Q^\ast l\cdot s_{\mathrm{min}}}{N+Q^\ast l\cdot s_{\mathrm{min}}})\approx 1.999992
\end{equation}

\noindent (3) By estimating the cost of replacing a 128-bit QKD-key, we can calculate the security benefits gained from using QKD-keys:
\begin{equation}\label{Formula31}
\frac{Q^\ast\times [1+\log_{2}(1+\frac{Q^\ast l\cdot s_{\mathrm{min}}}{N+Q^\ast\cdot s_{\mathrm{min}}})]}{k\times C_{\mathrm{QKD\text{-}key}}}=\frac{123577\times 1.999992}{2\times  C_{\mathrm{QKD\text{-}key}}}
\end{equation}

\noindent As shown in Fig.\ref{CBC-key},  assuming the cost of a single QKD-key is $1$, in CBC mode, increasing the frequency of rekeying (i.e.,shortening the QKD-key renewal period) leads to an increase in security strength but a decrease in the overall security benefits.

\noindent\textbf{ECBC-MAC mode calculation}

\begin{figure*}[htbp]
\centering
\includegraphics[width=0.8\textwidth]{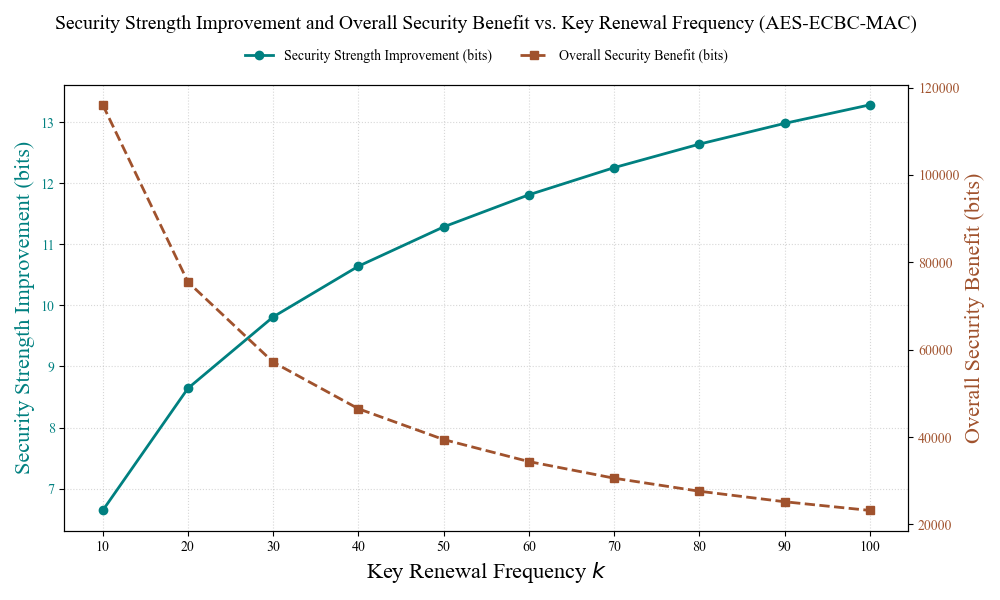} 
\caption{Relationship between security improvement and QKD-key renewal period in ECBC-MAC mode}
\label{CBC-MAC-key}
\end{figure*}

\noindent (1) Calculate the maximum QKD-key renewal period $Q^\ast$, by solving the inequality:
\begin{equation}\label{Formula32}
2\frac{96\times Q^\ast}{2^{126}}+\frac{96^2\times (Q^\ast)^2+(Q^\ast)^2+2}{2^{128}}\leq \frac{1}{2^{80}}
\end{equation}
\noindent We obtain $Q^\ast=174739$. The maximum encrypted data volume is $174739\times 1.5\mathrm{KB}=262109\mathrm{KB}\approx 256\mathrm{MB}$. Therefore, the key should be renewed after each QKD-key encrypts 256 MB of data.

\noindent (2) Assume $k=2$, meaning $2$ uniform QKD-key renewals are executed (i.e., the number of files encrypted by a single QKD-key is reduced to $Q^\ast/2$). The security strength improvement is:
\begin{equation}\label{Formula33}
1+\log_{2}(1+\frac{Q^{\ast ^2}(l^2+1)-4}{\frac{8NQ^\ast l}{s_{\mathrm{min}}}+Q^{\ast ^2}(l^2+1)+8})\approx 1.99996
\end{equation}

\noindent (3) By estimating the cost of replacing a 128-bit QKD-key, we can calculate the security benefits gained from using QKD-keys:
\begin{equation}\label{Formula34}
\begin{aligned}
\frac{
Q^\ast
\Bigl[
1+\log_{2}\!\Bigl(
1+
\frac{
{Q^\ast}^{2}(l^{2}+1)-4
}{
\tfrac{8NQ^\ast l}{s_{\mathrm{min}}}
+{Q^\ast}^{2}(l^{2}+1)+8
}
\Bigr)
\Bigr]
}{
k \times C_{\mathrm{QKD\text{-}key}}
}
&= \\
&\frac{174739 \times 1.99996}
{2 \times C_{\mathrm{QKD\text{-}key}}}
\end{aligned}
\end{equation}

\noindent As shown in Fig.\ref{CBC-MAC-key},  assuming the cost of a single QKD-key is $1$, in ECBC-MAC mode, increasing the frequency of rekeying (i.e., shortening the QKD-key renewal period) leads to an increase in security strength but a decrease in the overall security benefits.

\section{Conclusion}
This paper addresses the lack of precise quantification regarding security reduction caused by key reuse when integrating QKD-keys with classical block ciphers. We establish a systematic and precise model for computing the QKD-key renewal period, together with a quantitative method for evaluating the corresponding security benefits. Focusing on three representative block cipher modes—CTR, CBC, and ECBC-MAC—we have developed a concrete security analysis framework that yields the maximum number of files $Q^\ast$ that can be safely encrypted using a single QKD-key for a given security target. This provides a precise theoretical basis for QKD-key renewal to replace the rough estimation of the traditional $2^{n/2}$ boundary.

In addition, this paper organically combines the \\information-theoretic security characteristics of QKD-keys with the computational security of block ciphers, supporting the efficient and secure integration of block cipher applications in QKD networks. The proposed benefit-cost evaluation formula offers decision support for engineering deployment. Finally, specific parameter settings are provided for reference (to be adapted to the target 
scenario), along with examples of how to apply the method.

\begin{acknowledgements}
This research was primarily funded by the "Soft Science Research" project under the 2025 High-Level Institutional Development and Operation Program of Shanghai, People's Republic of China (No. 25692107800), the Guangdong Provincial Key Area R\&D Program (No. 2020B03\\03010001), the Quantum Science and Technology-National Science and Technology Major Project (No. 2021ZD0301300), the National Wide-Area Quantum Secure Communication Backbone Network Project, and the "Eastern Data, Western Computing" Demonstration Project: Quantum Trusted Cloud Project.  We express our sincere gratitude for the support and assistance from these project funds.
\end{acknowledgements}



\appendix
\onecolumn
\section{Formula Derivation}
This appendix provides the derivations for Equations
(\ref{Formula11}), 
(\ref{Formula17}), and (\ref{Formula23}).


(\ref{Formula11}):
\begin{equation}
\begin{split}
& -\log_{2}\!\left(
\frac{(Q^\ast/k)l}{s_{\min}}
+\frac{2(Q^\ast/k)^2l}{N}
\right)
-\left[
-\log_{2}\!\left(
\frac{Q^\ast l}{s_{\min}}
+\frac{2(Q^\ast)^2l}{N}
\right)
\right] \\
=& \log_{2}\!\left(
\frac{Q^\ast l}{s_{\min}}
+\frac{2(Q^\ast)^2l}{N}
\right)
-\log_{2}\!\left(
\frac{Q^\ast l}{ks_{\min}}
+\frac{2(Q^\ast)^2l}{k^2N}
\right) \\
=& \log_{2}\!\left(
\frac{
\frac{Q^\ast l}{s_{\min}}
+\frac{2(Q^\ast)^2l}{N}
}{
\frac{Q^\ast l}{ks_{\min}}
+\frac{2(Q^\ast)^2l}{k^2N}
}
\right) \\
=& \log_{2}\!\left(
\frac{
1+\frac{2Q^\ast s_{\min}}{N}
}{
\frac{1}{k}+\frac{2Q^\ast s_{\min}}{k^2N}
}
\right) \\
=& \log_{2}\!\left(
k\cdot
\frac{
1+\frac{2Q^\ast s_{\min}}{N}
}{
1+\frac{2Q^\ast s_{\min}}{kN}
}
\right) \\
=& \log_{2}k
+\log_{2}\!\left(
1+\frac{2(k-1)Q^\ast s_{\min}}{kN+2Q^\ast s_{\min}}
\right).
\nonumber
\end{split}
\end{equation}


(\ref{Formula17}):
\begin{equation}
\begin{split}
& -\log_{2}\!\left(
\frac{(Q^\ast/k)l}{s_{\min}}
+\frac{2(Q^\ast/k)^2l^2}{N}
\right)
-\left[
-\log_{2}\!\left(
\frac{Q^\ast l}{s_{\min}}
+\frac{2(Q^\ast)^2l^2}{N}
\right)
\right] \\
=& \log_{2}\!\left(
\frac{Q^\ast l}{s_{\min}}
+\frac{2(Q^\ast)^2l^2}{N}
\right)
-\log_{2}\!\left(
\frac{Q^\ast l}{ks_{\min}}
+\frac{2(Q^\ast)^2l^2}{k^2N}
\right) \\
=& \log_{2}\!\left(
\frac{
\frac{Q^\ast l}{s_{\min}}
+\frac{2(Q^\ast)^2l^2}{N}
}{
\frac{Q^\ast l}{ks_{\min}}
+\frac{2(Q^\ast)^2l^2}{k^2N}
}
\right) \\
=& \log_{2}\!\left(
\frac{
1+\frac{2Q^\ast l s_{\min}}{N}
}{
\frac{1}{k}+\frac{2Q^\ast l s_{\min}}{k^2N}
}
\right) \\
=& \log_{2}\!\left(
k\cdot
\frac{
1+\frac{2Q^\ast l s_{\min}}{N}
}{
1+\frac{2Q^\ast l s_{\min}}{kN}
}
\right) \\
=& \log_{2}k
+\log_{2}\!\left(
1+\frac{2(k-1)Q^\ast l s_{\min}}{kN+2Q^\ast l s_{\min}}
\right).
\nonumber
\end{split}
\end{equation}

(\ref{Formula23}):
\begin{equation}
\begin{split}
&{-}\log_{2}\!\left(
2\frac{(Q^{\ast}/k)\,l}{s_{\min}}
+\frac{(Q^{\ast}/k)^{2}l^{2}+(Q^{\ast}/k)^{2}+2}{2N}
\right)
-\left[
-\log_{2}\!\left(
2\frac{Q^{\ast}l}{s_{\min}}
+\frac{Q^{\ast 2}l^{2}+Q^{\ast 2}+2}{2N}
\right)
\right] \\[4pt]
&=\log_{2}\!\left(
\frac{
2\,\dfrac{Q^{\ast}l}{s_{\min}}
+\dfrac{Q^{\ast 2}(l^{2}+1)+2}{2N}
}{
\dfrac{2Q^{\ast}l}{ks_{\min}}
+\dfrac{Q^{\ast 2}l^{2}/k^{2}+Q^{\ast 2}/k^{2}+2}{2N}
}
\right) \\[4pt]
&=\log_{2}\!\left(
\frac{
4NQ^{\ast}l+s_{\min}\bigl[Q^{\ast 2}(l^{2}+1)+2\bigr]
}{
\dfrac{4NQ^{\ast}l}{k}
+s_{\min}\bigl[\dfrac{Q^{\ast 2}l^{2}}{k^{2}}+\dfrac{Q^{\ast 2}}{k^{2}}+2\bigr]
}
\right) \\[4pt]
&=\log_{2}\!\left(
\frac{
k\Bigl(4NQ^{\ast}l+s_{\min}\bigl[Q^{\ast 2}(l^{2}+1)+2\bigr]\Bigr)
}{
4NQ^{\ast}l
+s_{\min}\bigl[\dfrac{Q^{\ast 2}l^{2}}{k}+\dfrac{kQ^{\ast 2}}{k^{2}}+2k\bigr]
}
\right) \\[4pt]
&=\log_{2}k
+\log_{2}\!\left(
\frac{
4NQ^{\ast}l+s_{\min}\bigl[Q^{\ast 2}(l^{2}+1)+2\bigr]
}{
4NQ^{\ast}l
+s_{\min}\bigl[\dfrac{Q^{\ast 2}l^{2}}{k}+\dfrac{kQ^{\ast 2}}{k^{2}}+2k\bigr]
}
\right) \\[4pt]
&=\log_{2}k
+\log_{2}\!\left(
1+
\frac{
s_{\min}\Bigl[
Q^{\ast 2}l^{2}\!\left(1-\tfrac{1}{k}\right)
+Q^{\ast 2}\!\left(1-\tfrac{1}{k}\right)
+2(1-k)
\Bigr]
}{
4NQ^{\ast}l
+s_{\min}\bigl[\dfrac{Q^{\ast 2}l^{2}}{k}+\dfrac{kQ^{\ast 2}}{k^{2}}+2k\bigr]
}
\right) \\[4pt]
&=\log_{2}k
+\log_{2}\!\left(
1+
\frac{
Q^{\ast 2}(l^{2}+1)(1-1/k)+2(1-k)
}{
4N\dfrac{Q^{\ast}l}{s_{\min}}
+Q^{\ast 2}(l^{2}+1)/k+2k
}
\right).
\nonumber
\end{split}
\end{equation}

\end{document}